\documentclass[twocolumn,aps,prb,showpacs,superscriptaddress]{revtex4-1}
\usepackage[latin9]{inputenc}
\usepackage{amsmath}
\usepackage{amssymb}
\usepackage{graphicx}

\makeatletter
\usepackage{color}
\usepackage{comment}

\def\XXint#1#2#3{{\setbox0=\hbox{$#1{#2#3}{\int}$ }
		\vcenter{\hbox{$#2#3$ }}\kern-.5\wd0}}

\makeatother

\begin{document}
	
	\title{Generation of Teraherz Oscillations by Thin Superconducting Film
		in Fluctuation Regime}
	
	\author{T.~M.~Mishonov}
	\email[E-mail: ]{mishonov@gmail.com}
	
	
	\affiliation{Department of Theoretical Physics, Faculty of Physics, St.~Clement
		of Ohrid University at Sofia, 5 James Bourchier Blvd., BG-1164 Sofia,
		Bulgaria}
	
	\author{A.~M.~Varonov}
	\email[E-mail: ]{avaronov@phys.uni-sofia.bg}
	
	
	\affiliation{Department of Theoretical Physics, Faculty of Physics, St.~Clement
		of Ohrid University at Sofia, 5 James Bourchier Blvd., BG-1164 Sofia,
		Bulgaria}
	
	\author{I.~Chikina}
	
	\affiliation{LIONS, NIMBE, CEA, CNRS, Universit{\`a} Paris-Saclay, CEA Saclay
		91191 Gif sur Yvette Cedex, France}
	
	\author{A.~A.~Varlamov}
	
	\affiliation{CNR-SPIN, Viale del Politecnico 1, I-00133, Rome, Italy}
	
	\begin{abstract}
Explicit analytical expressions for conductivity
	 of a superconducting film above and below critical temperature
		in an arbitrary electric field are derived in the frameworks of the
		time dependent Ginzburg-Landau theory. It is confirmed that slightly
		below critical temperature the differential conductivity of superconducting
		film can become negative for small enough values of electric field.
		This fact may cause generation of electromagnetic oscillations if
		the superconducting film is appropriately coupled of with a resonator.
		Their maximal frequency is proportional to the value of critical temperature
		of superconducting transition. The obtained results can stimulate
		the development of Terahertz generators on the basis of high temperature
		superconducting films. 
	\end{abstract}
	
	\date{January 28, 2019}
	
	\maketitle

\section{Introduction}

\label{Sec:Introduction} Half a century ago Churilov, Dmitriev and
Beskorsiy\cite{Churilov:69} observed generation of high-frequency
monochromatic oscillations by thin (25~nm) superconducting tin film
included in a resonance circuit and being in resistive state. Analyzing
this phenomenon Gor'kov\cite{Gorkov:70} became interested in the
fact that such film exposed to electric field at temperatures slightly
below the critical one, remains stable against the occurrence of infinitesimal
nuclei of the superconducting phase down to the very weak fields $E$.
He microscopically derived the corresponding current-voltage characteristics
$j(E)$ accounting for supercurrent caused by the order parameter
fluctuations. It was turned out that close to critical temperature
the electric field can break down yet fragile Cooper pairs, which results
in suppression of the superconducting current component.
Above $T_{c0}$ (the critical temperature of transition in absence of electric field)
this electric field sensitivity of fluctuation Cooper pairs results
in suppression of the positive Aslamazov-Larkin fluctuation contribution
to the Drude conductivity.\cite{H69,RV} In contrast, below $T_{c0}$,
the correction related to breaking of the true Cooper pairs by electric
field changes sign and, consequently, leads to appearance of the negative
differential conductivity.

Such fall down of the differential conductivity of superconducting
film is the precursor of radiation generation. It turns out that the
frequency of such radiation depends on the closeness of the film temperature
to the critical one $T_{c0}$. The maximal value of the frequency
of generated radiation can be reached using the high temperature superconducting
films ($\nu\propto5\,$ THz for the film with $T_{c0}=90$~K), i.e.,
it falls into the Terahertz region which is now an intensive field
of research. This circumstance revives the interest in the cited above
two half-century old papers, whose re-reading opens the exciting perspective
to develop the new type of THz generators based on nanoscale hybrid
superconducting devices super-cooled in the normal state by small
electric field. The purpose of this communication is to present the
analytical formulas for differential conductivity, which can be useful
for technical development of such devices  and to determine the threshold field of generation.

\section{The effect of electric field on Cooper pairs}

\label{Sec:Theory}

The break of fluctuation Cooper pair by electric field above $T_{c0}$
can be understood qualitatively as follows. The electrons correlated
in Cooper pair have almost the opposite momenta. Therefore, the same
acceleration which each of them acquires in electric field $E$ results
in the growth of velocity for one of them and decrease for another.
This, in its turn, leads to the increase of the distance between electrons.
The pair decays if this distance reached during the pair lifetime
$\tau_{\mathrm{_{GL}}}=\pi\hbar/8k_{\mathrm{_{B}}}(T-T_{c0})$ exceeds
the coherence length $\xi(\epsilon)\sim\xi/\sqrt{|\epsilon|}$, where
$\epsilon=\left(T-T_{c0}\right)/T_{c0}$ is the reduced temperature
(which can acquire both positive and negative values) and the parameter
of Ginzburg-Landau theory $\xi$ will be defined below. In other words,
starting from some characteristic, temperature dependent, value of
the intensity of electric field $E_{c}^{\left(+\right)}(\epsilon)$,
the electrons acceleration  becomes so large, that at the distance of the order $\xi(\epsilon)$ the electrons change their energy by the value of the order of $T-T_{c0}$ corresponding to the fluctuation Cooper pair ``binding energy''.  The described mechanism results in the additional  with respect to thermal, field
depending, decay of fluctuation pairs and respective deviation of
the voltage-current characteristics from the Ohm law. Below $T_{c0}$
the mechanism of suppression of superconductivity is similar: the
electric field breaks up potentially emerging, still weak, Cooper
pairs and does not allow superconducting state to be established.

One can see that the threshold electric field $E_{c}^{\left(+\right)}(\epsilon)$,
where the nonlinear effects begin to manifest themselves above critical
temperature is determined from the condition $eE_{c}^{\left(+\right)}(\epsilon)\centerdot\xi(\epsilon)\propto k_{\mathrm{_{B}}}\left(T-T_{c0}\right)$
and it tends zero as $\epsilon{}^{3/2}$ when temperature verges towards
$T_{c0}$ \cite{H69}
\begin{equation}
E_{c}^{\left(+\right)}(\epsilon)=\frac{16\sqrt{3}\:k_{\mathrm{_{B}}}T_{c0}}{\pi e\xi}|\epsilon|^{3/2},\quad\epsilon>0.\label{Ec+}
\end{equation}
Here $\xi$ is the parameter of Ginzburg-Landau theory calculated
by Gor'kov \cite{G59}: 
\begin{equation}
\xi^{2}\!=\!-\frac{v_{F}^{2}\tau^{2}}{3}\!\left[\psi\left(\!\frac{1}{2}\!+\!\frac{1}{4\pi T\tau/\hbar}\right)\!-\!\psi\left(\frac{1}{2}\right)\!-\!\frac{\pi}{8T\tau/\hbar}\right].\label{xi}
\end{equation}
The latter can be experimentally determined from the slope of linear
extrapolation of the second critical field
\begin{equation}
-T_{c0}\left.\frac{\mathrm{d}H_{c2}}{\mathrm{d}T}\right|_{T_{c0}}=\frac{\Phi_{0}}{2\pi\xi^{2}},\label{hc2}
\end{equation}
with the magnetic flux $\Phi_{0}=\frac{\pi\hbar}{|e|}$.

While above $T_{c0}$ the applicability of TDGL equation is not questionable,
below the critical temperature TDGL theory is applicable only in the
case when a gap in quasi-particle spectrum is suppressed, for instance
by paramagnetic impurities\cite{EG68}. In the latter case the spin-flip
scattering time of electrons forming a Cooper pair $\tau_{s}$ should
be of the order of its its inverse condensation energy: $\tau_{s}\sim\Delta^{-1}$
(see Refs.~[\onlinecite{AG60,M68}]). It is clear that in the case under
consideration (close to $T_{c0}$) strong electric field will impede
to establishment of the superconducting state until its work performed
on the electrons ``forming'' Cooper pair at the distance of the
order of correlation length remains larger than the value of gap:
$eE_{c}^{\left(-\right)}(|\epsilon|)\centerdot\xi(|\epsilon|))\thicksim\Delta\left(\epsilon\right)$.
This condition results in the linear dependence of the edge of nonlinearities
on reduced temperature:
\begin{equation}
E_{c}^{\left(-\right)}\left(\epsilon\right)\sim\frac{\Delta_{BCS}}{e\xi}|\epsilon| \label{Ec-}
\end{equation}
with $ \Delta_{BCS}$  as the BCS value of the superconducting gap at zero temperature. Comparison of this equation, written in the form $\left[eE_{c}^{\left(-\right)}(|\epsilon|)\centerdot\xi(|\epsilon|))\right]^{-1}\Delta\left(\epsilon\right)\thicksim1$,
to the standard criterion of the gapless superconductivity $ \tau_{s}\Delta\left(\epsilon\right)\sim1 $
gives us the expression for the phase-breaking time arising due to
the presence of the electric field

\begin{equation}
\tau_{s}^{E}\left(\epsilon\right)=\left[eE_{c}^{*}(|\epsilon|)\centerdot\xi(|\epsilon|))\right]^{-1}=\frac{\Delta_{BCS}^{-1}}{\sqrt{|\epsilon|}}.\label{tauE}
\end{equation}

The general expression for the nonlinear current accounting for the
fluctuation conductivity in the case of a two dimensional (2D) superconductor
can be obtained in the vicinity of critical temperature in different
ways: using Boltzmann transport equation for fluctuation Cooper pairs
above $T_{c0}$, in the frameworks of the TDGL formalism, and in diagrammatic
approach.\cite{Gorkov:70,H69,Kulik:71,RV,Boltzman_FCP}
We will take it in the form, valid both below and above critical temperature
\begin{equation}
j(E)=\left[\sigma_{\mathrm{Dr}}+\overline{\sigma}_{\mathrm{_{AL}}}(|\epsilon|)\mathcal{\varXi}_{\pm}(E)\right] E, \label{totcurrent}
\end{equation}
with 
\begin{equation}
\mathcal{\varXi}_{\pm}(E)=\int_{0}^{\infty}\!\!\exp\left\lbrace -v\cdot\mathrm{sign\left(\epsilon\right)}-\!\alpha_{\pm}^{2}(E,\epsilon)\,v^{3}/3\right\rbrace \!\mathrm{d}v.\label{varXi}
\end{equation}
The sign of linear term in the exponent is determined by the sign
of reduced temperature $\epsilon$, i.e. above or below superconducting
transition the system stays. The parameter 
\begin{equation}
\alpha_{\pm}(E,\epsilon)=\frac{\sqrt{3}E}{E_{c}^{\left(\pm\right)}(|\epsilon|)}\label{alpha}
\end{equation}
is the dimensionless electric field normalized on the introduced above
$E_{c}^{\left(\pm\right)}(|\epsilon|)$, while the value 
\begin{equation}
\overline{\sigma}_{\mathrm{_{AL}}}(\epsilon)=\frac{e^{2}}{16\hbar|\epsilon|}
\end{equation}
above $T_{c0}$ has the sense of the two-dimensional Aslamazov-Larkin
conductivity and it determines the magnitude of the fluctuation effect.

The variable of integration in Eq.~(\ref{totcurrent}) can be also
interpreted in physical terms. This is nothing else as the dimensionless
time: 
\begin{equation}
v=\frac{t}{\tau_{\mathrm{_{GL}}}/2},\quad\tau_{\mathrm{_{GL}}}=\frac{\pi\hbar}{8k_{\mathrm{_{B}}}T_{c0}|\epsilon|}.
\end{equation}
It is normalized by the half of Ginzburg-Landau decay time $\tau_{\mathrm{_{GL}}}$
of the order parameter, extended symmetrically with respect to critical
temperature for the temperatures below the latter.

For intuitive interpretation it is convenient to introduce the decay
rate of fluctuation Cooper pairs 
\begin{equation}
\nu(\epsilon)=\frac{2}{\tau_{\mathrm{_{GL}}}}=\frac{16k_{\mathrm{_{B}}}T_{c0}\epsilon}{\pi\hbar}.\label{decay_rate}
\end{equation}
The multiplier $\exp\left[-{\mathrm{sign}}(\epsilon)\cdot v\right]=\exp(-{\mathrm{sign}}\left(\epsilon\right)\nu t)$
in the integrand in Eq.~(\ref{totcurrent}) above $T_{c0}$ describes
the spontaneous exponential decay of fluctuation Cooper pairs. Close
to critical temperature $T\rightarrow T_{c0}$ one can see the critical
slowing down of this process: $\nu\rightarrow0$. Below the critical
temperature the sign of the first term in the exponent of Eq. (\ref{varXi})
changes, what, formally, corresponds to the lasing of fluctuation
Cooper pairs instead of their decay (i.e. the exponential increment
of their concentration, as the light intensity in lasing media). Yet,
this lasing is restricted by the second term in the exponent of Eq.~(\ref{varXi}), which dominates on the first one when $v\gtrsim1/\alpha_{-}(\epsilon)$.
\begin{figure}[tbh]
	\includegraphics[ width=1.0 \columnwidth ]{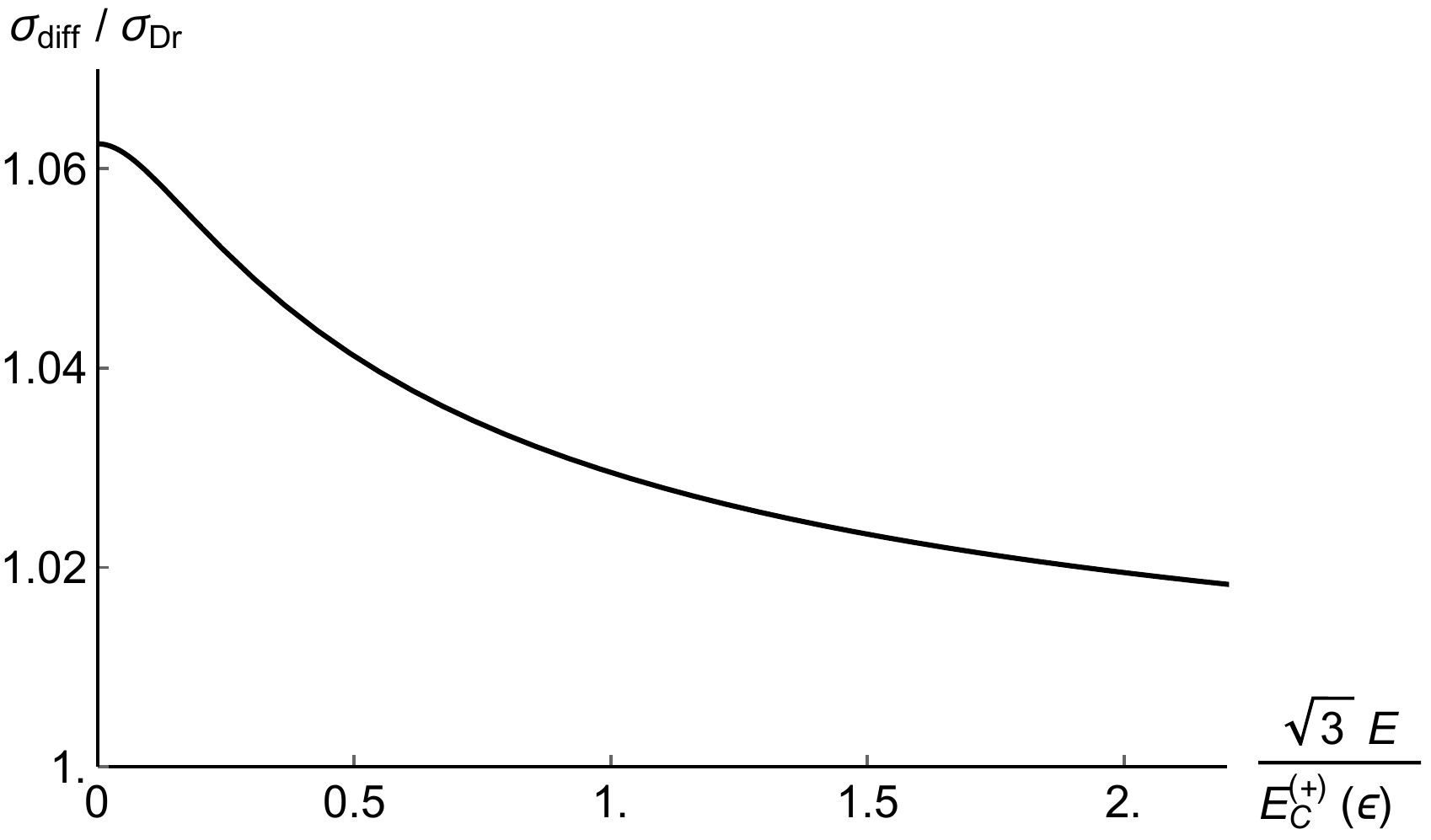}
	\caption{Differential conductivity as a function of the dimensionless electric field above the critical temperature ($\epsilon =0,01$, $\sigma_{\mathrm{Dr}}=100e^2/\hbar$).}
	\label{fig:j_above} 
\end{figure}

As it was mentioned above, the arising close to critical temperature
Cooper pairs decay in the presence of electric field is due to the increase
of their kinetic energy. For small enough electric fields, however,
the decay is delayed and the fluctuation conductivity can reach significant
values and even can dominate over the normal Drude conductivity. In
short, above $T_{c0}$ the first term in the exponent of integrand
in Eq.~(\ref{totcurrent}) is negative, while below it becomes positive.
This difference results in the qualitatively different manifestations
of fluctuations in nonlinear conductivity.

The effect of electric field on fluctuation Cooper pairs is accounted
for by the cubic term in the exponent of Eq.~(\ref{varXi}). It becomes
strong enough when the kinetic energy acquired due to acceleration
in electric field exceeds the GL ``binding energy'', what
happens when $E\sim E_{c}$. In this region of fields fluctuation
Cooper pairs decay and the corresponding  contribution
to the total current with the further growth of the field intensity
decreases.

Let us analyze Eqs. (\ref{totcurrent})-(\ref{varXi}) separately
above and below the critical temperature.
\begin{figure}[tbh]
	\begin{centering}
		\includegraphics[width=1.0\columnwidth]{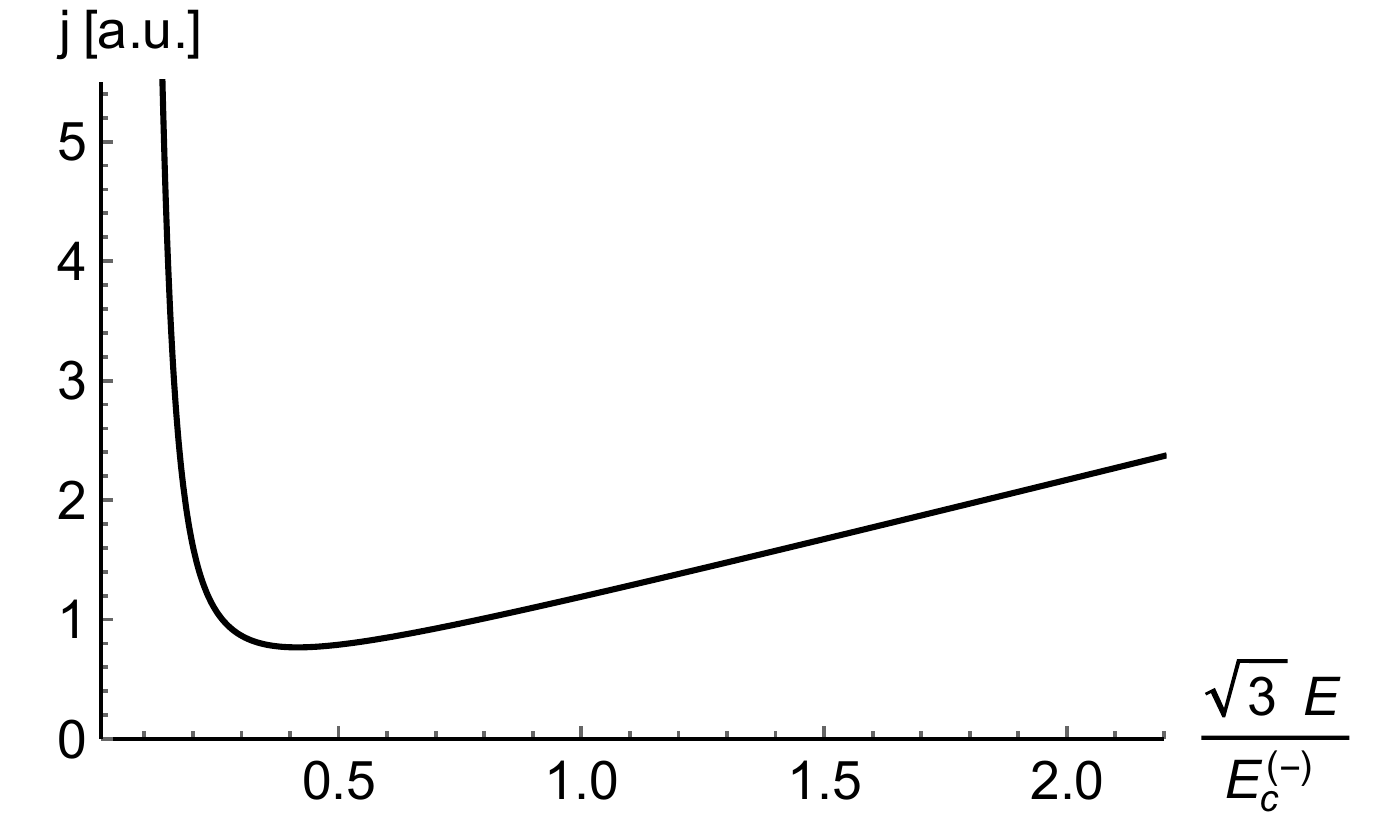} 
		\par\end{centering}
	\caption{ Current density $j$ (arbitrary units) in a film ``super-cooled'' by electric field versus the value of the latter below the critical temperature ($\epsilon =-0,01$, $\sigma_{\mathrm{Dr}}=100e^2/\hbar$) (compare with the schematic
		figure of Gor'kov.~\cite{Gorkov:70})}	
	\label{fig:current} 
\end{figure}

\subsubsection{Above critical temperature}

The integration in Eq.~(\ref{varXi}) above the transition temperature
can be performed exactly in terms of the Bessel and hypergeometric
functions. The corresponding expression for $\mathcal{\varXi}_{+}$
acquires the form: 
\begin{widetext}
\begin{eqnarray}
\mathcal{\varXi}_{+}(\alpha_{+})=\frac{2\pi}{3\sqrt{3}\alpha_{+}}\left\lbrace \mathrm{J}_{-\frac{1}{3}}\left(\frac{2}{3\alpha_{+}}\right)-\mathrm{J}_{\frac{1}{3}}\left(\frac{2}{3\alpha_{+}}\right)\right\rbrace +\frac{1}{2\alpha_{+}^{2}}\,_{1}\mathrm{F}_{2}\left(1;\frac{4}{3},\frac{5}{3};-\frac{1}{9\alpha_{+}^{2}}\right) & \approx & \begin{cases}
1,\qquad\alpha_{+}\ll1,\\
\Gamma\left(\frac{4}{3}\right)\left(\frac{3}{\alpha_{+}^{2}}\right)^{1/3},\quad\alpha_{+}\gg1.
\end{cases}\label{cond_approx}
\end{eqnarray}
\end{widetext}

Comparing Eqs. (\ref{totcurrent}) and (\ref{cond_approx}) one can
see that in the region of small fields the fluctuation correction
to the conductivity is positive and equal to the Aslamazov-Larkin paraconductivity.
The corresponding correction monotonously decreases as $\left(E_{c}^{\left(+\right)}(|\epsilon|)/E\right)^{2/3}$
(compare to Ref.~[\onlinecite{RV}]) when the fields exceed the threshold
value. The corresponding field dependence of differential conductivity for reasonable value
$ \sigma_{\mathrm{Dr}}=100e^{2}/\hbar$ is shown in Fig. \ref{fig:j_above}.

\subsubsection{Below critical temperature}

Below the critical temperature the behavior of the integrand function
in Eq. (\ref{varXi}) strikingly  differs from that one in the
previous subsection due to the growth of the linear term (instead
of its decrease) in the exponent. Yet, the integral still can be carried
out exactly: 
\begin{widetext}
\begin{eqnarray}
\mathcal{\varXi}_{-}(\alpha_{-})=\frac{2\pi}{3\sqrt{3}\alpha_{-}}\left\lbrace \mathrm{I}_{-\frac{1}{3}}\left(\frac{2}{3\alpha_{-}}\right)+\mathrm{I}_{\frac{1}{3}}\left(\frac{2}{3\alpha_{-}}\right)\right\rbrace +\frac{1}{2\alpha_{-}^{2}}\,_{1}\mathrm{F}_{2}\left(1;\frac{4}{3},\frac{5}{3};\frac{1}{9\alpha_{-}^{2}}\right)\approx\begin{cases}
\sqrt{\frac{\pi}{\alpha_{-}}}\exp\left(\frac{2}{3\alpha_{-}}\right)\gg1,\alpha_{-}\ll1,\\
\Gamma\left(\frac{4}{3}\right)\left(\frac{3}{\alpha_{-}^{2}}\right)^{1/3},\qquad\alpha_{-}\gg1.
\end{cases}\label{cond_approx-1}
\label{Xi-}
\end{eqnarray}
\end{widetext}
Corresponding behavior of current as the function of the electric field
below critical temperature is illustrated in Fig.~\ref{fig:current}.

As was already explained above, the superconducting transition of thin film
subjected of electric field is delayed to lower temperatures. Nevertheless,
the current in this super-cooled state formally growths when the electric field
decreases below some critical value $\alpha_{osc}$ corresponding to the minimum in Fig. \ref{fig:current}.
In order to determine the latter, one can calculate the differential conductivity
\begin{equation}
\sigma_{\mathrm{diff}}=\mathrm{d}j(E)/\mathrm{d}E=\sigma_{\mathrm{Dr}}+\sigma_{\mathrm{_{AL}}}(|\epsilon|)\mathcal{F}_{-}(\alpha_{-}),\label{sigmadiff}
\end{equation}
where the function $\mathcal{F}_{-}$ is obtained by  differentiation of the current (see Eq.~(\ref{totcurrent})) with $\mathcal{\varXi}_{-}(\alpha_{-})$ taken from Eq.~(\ref{Xi-}).
It  can be expressed exactly in terms of the modified Bessel functions and the hypergeometric, function
\begin{figure}[b]
	\includegraphics[width=1\columnwidth]{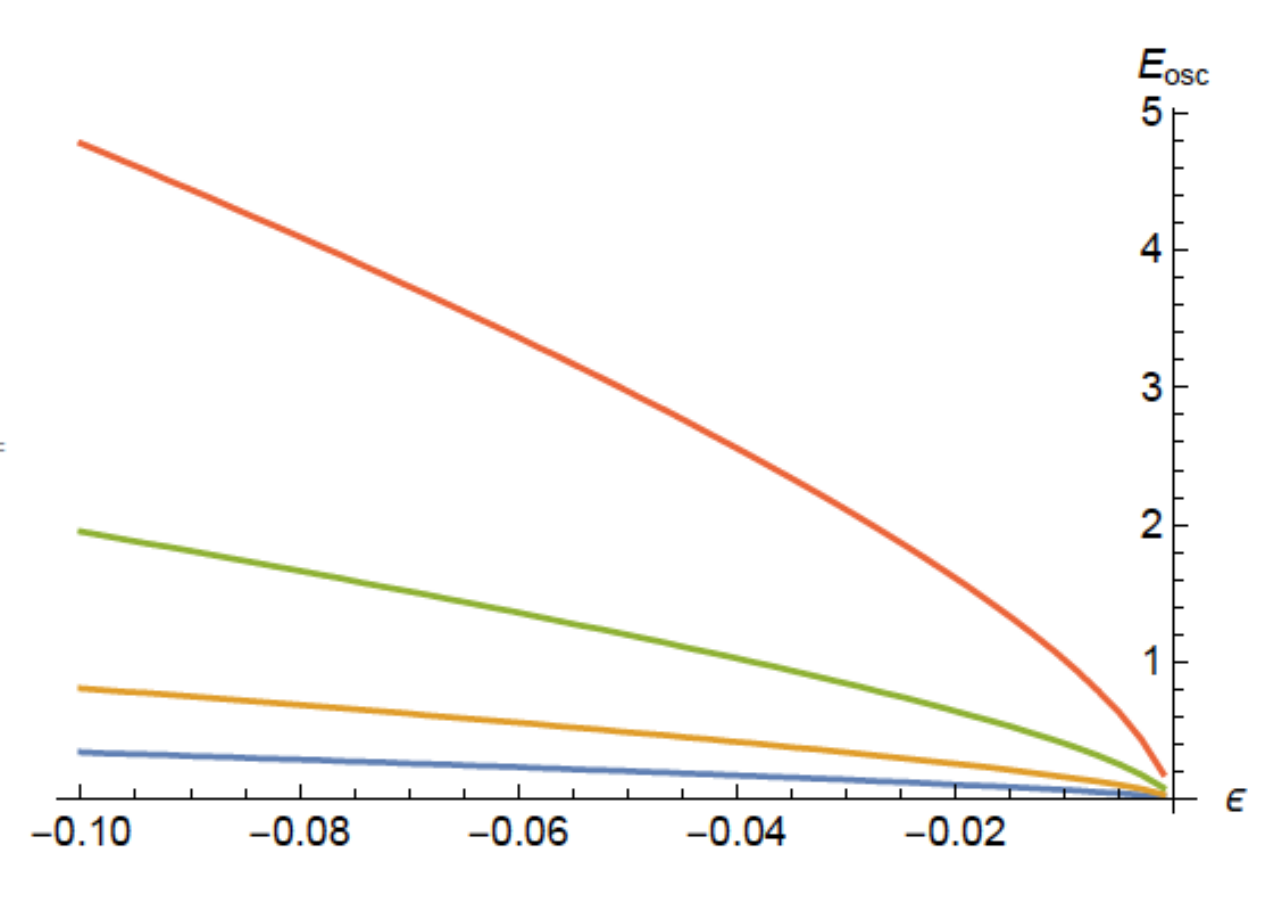}
	\caption{The threshold electric field $E_{\mathrm{osc}}$ of the oscillations generation (arbitrary units) 	as the function of reduced temperature $\epsilon$ for four values of the film conductivity (going from the bottom up): $\sigma_{\mathrm{Dr}}=50e^{2}/\hbar$, $\sigma_{\mathrm{Dr}}=25e^{2}/\hbar$, $\sigma_{\mathrm{Dr}}=12e^{2}/\hbar$, $\sigma_{\mathrm{Dr}}=6e^{2}/\hbar$.}
	\label{fig:osc} 
\end{figure}
\begin{widetext}
\begin{eqnarray}
\mathcal{F}_{-}(\alpha) & \equiv & -\int_{0}^{\infty}\left(2\alpha^{2}v^{3}/3-1\right)\exp(v-\alpha^{2}v^{3}/3)\,\mathrm{d}v \label{Ac}\\
 & = & \frac{2}{27\alpha^{2}}\!\left\lbrace 3|\alpha|\mathrm{K}_{-\frac{1}{3}}\left(\!\frac{2}{3|\alpha|}\!\right)\!-\!2\pi\sqrt{3}\left[\mathrm{I}_{\frac{4}{3}}\left(\!\frac{2}{3|\alpha|}\!\right)\!+\!\mathrm{I}_{\frac{2}{3}}\left(\!\frac{2}{3|\alpha|}\!\right)\right]\right\rbrace \!+\!\frac{1}{2\alpha^{2}}\,_{1}\mathrm{F}_{2}\left(\!1;\frac{4}{3},\frac{5}{3};\!\frac{1}{9\alpha^{2}}\!\right)\!-\!\frac{1}{\alpha^{2}}\,_{1}\mathrm{F}_{2}\left(\!2;\frac{4}{3},\frac{5}{3};\!\frac{1}{9\alpha^{2}}\!\right).\nonumber 
\end{eqnarray}
\end{widetext}
The  loss of stability of the system and, correspondingly, the condition for
the  possibility of electromagnetic oscillations generation at fixed temperature
 is determined by the requirement $\sigma_{\mathrm{diff}} (E_{\mathrm{osc}}, \epsilon<0)=0$,
which leads to the transcendental equation 
\begin{equation}
\mathcal{F}_{-}(\alpha_{\mathrm{osc}})=-\frac{\sigma_{\mathrm{_{Dr}}}}{e^{2}/16\hbar}|\epsilon|,\quad\epsilon<0.\label{eosc}
\end{equation}
The value of $\alpha_{\mathrm{osc}}$ evidently depends on the conductance of the film and closeness to the transition temperature. 

For the low Ohmic film ($\sigma_{\mathrm{_{Dr}}}\gg e^{2}/\hbar$) the function
$\mathcal{F(\alpha)}$ can be simply approximated in elementary functions
applying the steepest descend method for the integral (\ref{Ac}):
\begin{eqnarray}
\mathcal{F}_{-}(\alpha)  \approx-  \begin{cases}
\sqrt{\frac{\pi}{\alpha}}\left(\frac{2}{3\alpha}-1\right)\exp\left(\frac{2}{3\alpha}\right), \quad \alpha\ll1,\\ \\
\frac{2\Gamma\left(\frac{4}{3}\right)}{\left(3\alpha\right)^{2/3}},\qquad\alpha\gg1.\label{GK}
\end{cases}
\end{eqnarray}
Let us note that  the low Ohmic film approximation for the fluctuation induced current $\j_{\mathrm{fluct}}\propto E^{1/2}\exp(\mathrm{const}/E)$
was firstly pointed out by Gor'kov \cite{Gorkov:70} (compare with the upper line in Eq. (\ref{GK}) having in mind Eq. (\ref{totcurrent})).

The solution of Eq. (\ref{eosc}) in assumption that $\alpha_{\mathrm{osc}}\ll1$
(i.e. in the approximation of Eq. (\ref{GK})) with logarithmic accuracy
gives the value of critical electric field 
\begin{equation}
E_{\mathrm{osc}}=\frac{2E_{c}^{\left(-\right)}\left(|\epsilon|\right)}{3^{3/2}\ln\left(\dfrac{16\hbar\sigma_{\mathrm{Dr}}}{e^{2}}\right)}\ll E_{c}^{\left(-\right)}\left(|\epsilon|\right).
\end{equation}
One can see that in this approximation the threshold of instability increases linearly on temperature with moving away from $T_{c0}$
\begin{equation}
E_{\mathrm{osc}}(\epsilon <0)\sim\frac{\Delta_{BCS}}{e \xi}\frac1{ \ln\left(\dfrac{16\hbar\sigma_{\mathrm{Dr}}}{e^{2}}\right)}|\epsilon|.
\end{equation}

The dependencies of $E_{\mathrm{osc}}(\epsilon)$ for different values of $\sigma_{\mathrm{Dr}}$ obtained in result of numerical solution of Eq.~(\ref{eosc}) are presented in Fig. \ref{fig:osc}. One can see that in the limit of low-Ohmic film (the lowest curve) the threshold field indeed depends linearly on reduced temperature

\section{Conclusions}

\label{Sec:Conclusion} 

In this article we demonstrated that the negative differential conductivity of superconducting film at small electric fields in the vicinity of critical temperature  is
the ingredient for the loss of stability of the superconducting state and generation of electric
oscillations. This effect could be especially important in case of  high temperature superconductor films, where the
frequencies $k_{\mathrm{_{B}}}T_{c}/2\pi\hbar$ fall already in the
Teraherz region. The latter opens the perspectives for creation of a new type
generators of electric radiation. For every substrate it is necessary
to take into account the interface boundary condition, but the first step
of the technical applications will be the observation of the critical
point at which the differential conductivity is annulled and the system
losses its stability. The illustrative description of possible electronic
circuits will be described elsewhere.

\begin{acknowledgements} One of the authors (TMM) is grateful to
Damian Damianov, Evgeni Penev, Ana Posazhennikova, Mihail Mishonov,
and Yana Maneva for the collaboration at the early stages of the present
research; he appreciates stimulative help by Valya Mishonova. The
work was partially supported by COST action CA~16218 NANOCOHYBRI.
A.~A.~V. acknowledges EC for the RISE Project CoExAN GA644076.
\end{acknowledgements}




\begin{thebibliography}{1}
\bibitem{Churilov:69}G.~E.~Churilov, V.~M.~Dmitriev and A.~P.~Beskorsiy,
``Generation of high-frequency oscillations in thin superconducting
tin films'', ZhETF Pis. Red. \textbf{10}, 231-233 (1969).

\bibitem{Gorkov:70} L.~P.~Gorkov, ``Singularities of the resistive
state with current in thin superconducting films'', JETP Letters,
\textbf{11}, 32 (1970).

\bibitem{H69}J.~P.~Hurault, ``Nonlinear Effects on the Conductivity
of a Superconductor above Its Transition Temperature'', Phys. Rev.
179, 494 , 1969

\bibitem{Kulik:71} I.~O.~Kulik, ``Non-stationary Effects in the Resistive State of Superconducting Films'',  Sov. Phys. JETP \textbf{32}, 318 (1971).

\bibitem{RV} A.~A.~Varlamov and L.~Reggiani, ``Nonlinear fluctuation
conductivity of a layered superconductor: Crossover in strong electric
fields'', Phys. Rev. B \textbf{45}, 1060 (1992).

\bibitem{G59}L.~P.~Gorkov, ``Microscopic Derivation of the Ginzburg-Landau
Equations in the Theory of Superconductivity'', Sov. Phys. JETP \textbf{9}, 1364 (1959).

\bibitem{EG68} L.~P.~Gor'kov, G.~M.~Eliashberg, ``Generalization
of Ginzburg-Landau Equations for Non-Stationary Problems in the Case
of Alloys with Paramagnetic Impurities'', Sov. Phys. JETP \textbf{27}(2),
328-334 (1968).

\bibitem{AG60} A.~A.~Abrikosov, L.~P.~Gorkov, Sov. Phys. JETP
\textbf{12}, 1243 (1961).

\bibitem{M68} Kazumi Maki, ``Gapless Superconductivity'' in \textit{Superconductivity}, Vol.~1,
edited by R.~D.~Parks (Marcel Dekker Inc, New York, 1968), Chap.~18, pp.~1035--1106.

\bibitem{Boltzman_FCP} T.~M.~Mishonov and Y.G.~Maneva, arXiv:cond-mat/0604494;
T.~M.~Mishonov and M.~T.~Mishonov, arXiv:cond-mat/0505696; T.~Mishonov,
A.~Posazhennikova, J.~Indekeu, arXiv:cond-mat/0106168; T.~M.~Mishonov,
G.~V.~Pachov, I.~N.~Genchev, L.~A.~Atanasova, D.~Ch.~Damianov,
arXiv:cond-mat/0302046. 
\end{thebibliography}
\end{document}